\documentclass[aps,pra,twocolumn,showpacs,superscriptaddress]{revtex4-1}
\usepackage{graphics}
\usepackage{graphicx}
\usepackage{dcolumn}
\usepackage{bm}
\usepackage{color}
\usepackage{amssymb,amsmath}

\begin{document}

\title{Quasi-phase-matching $\chi^{(3)}-$parametric interactions in sinusoidally-tapered waveguides}

\author{Mohammed F. Saleh}

\affiliation{Institute of Photonics and Quantum Sciences, Heriot-Watt University, EH14 4AS Edinburgh, UK}
\affiliation{Department of Mathematics and Engineering Physics, Alexandria University, Alexandria, Egypt}

\begin{abstract}
In this article, I show how periodically-tapered waveguides can be employed as  efficient quasi-phase-matching schemes for  four-wave mixing parametric processes in third-order nonlinear materials.   As an example, a  thorough study of  enhancing third-harmonic generation in sinusoidally-tapered  fibres has been conducted. The quasi-phase-matching condition has been  obtained for nonlinear parametric interactions in these structures using Fourier-series analysis. The dependencies of the conversion efficiency of the third harmonic on the modulation amplitude, tapering period, longitudinal-propagation direction, and pump wavelength have been studied. In comparison to uniform waveguides, the conversion efficiency has been enhanced by orders of magnitudes. I envisage that this work can reshape the field of guided nonlinear optics using centrosymmetric materials.
\end{abstract}

\maketitle

\section{Introduction}
Optical   parametric nonlinear processes have been intensively harnessed in  applications such as optical-frequency conversion, amplification, and oscillation  \cite{Saleh07}. Efficient parametric wave-mixing  processes in nonlinear media are inherently constrained by energy and momentum conservation. The latter is known as the phase-matching condition. Two techniques have been successfully employed to satisfy this condition in second-order nonlinear media.  The first technique is based on interaction between non-collinear propagating photons in birefringent materials \cite{Kleinman62}.   In the second technique, known as quasi-phase-matching (QPM), the nonlinear coefficient $\chi^{(2)}$ is longitudinally  modulated with a specific period \cite{Armstrong62}. Using   periodically-poled ferroelectric crystals, where the nonlinear coefficient can be  flipped every half this period, this method has been experimentally demonstrated nearly after two decades from its proposal due to the lack of the suitable fabrication facilities at that time \cite{Feng80}. QPM structures have remarkably boosted the  classical and quantum optical nonlinear-frequency-conversion applications in bulk and integrated structures \cite{Fejer92,Hu13}. This technique enables, for instance: (i) Collinear interactions between co-polarised waves, results in exploiting the strongest component of the second-order nonlinear tensor. (ii) On-chip nonlinear interactions that lead to efficient and scalable optical integrated devices \cite{Tanzilli01}. (iii)  Tailoring the spectral properties of the output photons via engineering the poling pattern  \cite{Dosseva16}.  

In third-order  $\chi^{(3)}-$nonlinear media, few techniques have been used to satisfy the inherent difficult phase matching condition associated with four-wave mixing parametric processes. For instance: (i) Spontaneous  four-wave mixing in waveguides, in which two photons will coalesce to generate two other  photons, has been enabled via balancing the waveguide and material dispersion in narrow spectral range near the zero dispersion wavelength,   or  exploiting the nearly-phase matched  range, where the coherence length is much longer than the structure length \cite{Agrawal07,Sharping06,Fulconis07,Takesue08}. (ii) Third-harmonic generation has been enhanced via nonlinear  interactions in multimode waveguides \cite{Efimov03,Corona11,Moebius16}, using a hybrid photonic crystal fibre to allow interaction between a pump wave with its third-harmonic, both in the fundamental mode \cite{Cavanna16}, or exploiting slow-light effect introduced by photonic crystals \cite{Corcoran09}.

Having robust $\chi^{(3)}-$QPM platforms would push four-wave mixing parametric processes to a new regime, such as  on-demand spontaneous  four-wave mixing process, and efficient sum-frequency generation, where three different photons combine together to generate a photon at their sum frequency, with all the  photons in the fundamental mode. The inverse of the latter process, where one photon splits into three correlated or entangled photons `photon triplets', can  also take place. Having an efficient triplet source is beneficial for quantum-secret sharing applications \cite{Hillery99,Lance04}, as well as, building large-scale silicon-photonic quantum computers \cite{Rudolph16}. Few different QPM techniques in $\chi^{(3)}-$media  have been proposed and implemented for   parametric amplification,  third-harmonic generation, and controlling modulation instability,  such as cascaded  stages made  of a dispersion-shifted fibre and a single-mode uniform fibre   \cite{Kim01},   ultrasound waves in gas cells   \cite{Sapaev12}, counter-propagating train of pulses \cite{Jiang17}, and sinusoidally tapered fibres and silicon nanowires \cite{Tarnowski11,Driscoll12,Droques12,Droques13,Nodari15,Copie15}.

Periodically-tapered-waveguides  (PTWs) technique is a potential route to achieve robust $\chi^{(3)}-$QPM platforms, analogue to periodic poling, since by only longitudinally engineering the waveguide cross-section the phase-matching condition could be satisfied. This technique is currently hurdled by fabrication methods that limits the tapering period to the millimetre and sub-millimetre range in silicon nanowires \cite{Driscoll12}. Also, a careful analysis is essential to understand how simultaneous longitudinal variations of both linear and nonlinear properties of the waveguide modifies the well-known QPM condition in periodically-poled structures. In this article, I will explain using Fourier-series analysis how  periodically-tapered waveguides (PTWs) could act as an efficient analogue to QPM schemes in $\chi^{(3)}-$nonlinear media   for enhancing  third harmonic (TH) generation,  as an example of four-wave mixing parametric processes. The article is organised as follows: The model and the governing equations have been introduced in Sec. II. Sec. III discusses the mechanism of satisfying the phase matching condition in PTWs. Finally, Sec. IV is devoted for discussions and conclusions.

\section{Governing equations of third-harmonic generation in periodically-tapered waveguides}
Consider a TH parametric process in a single-mode PTW with a tapering period $\Lambda_\mathrm{T}$. A piecewise model has been developed to study this process in this type of waveguides, however, in general it could be applied for any other parametric processes. The procedure is the following: (i) Use Maxwell equations and assume that $\left|\partial_{z} \beta_{i}\right|  \ll \beta^2_{i}$, where $z$ is the longitudinal direction,  and $\beta_i$ is the propagation constant of a wave $i$.  (ii) Discretise the waveguide into infinitesimal segments, where the waveguide cross-section is assumed to be constant. An eigenvalue problem can be solved to determine the fundamental mode profile and its propagation constant. (iii) Apply the slowly-varying envelope approximation, the interaction between a fundamental pump wave and its TH  in the limit of CW-approximation inside each segment  is governed by \cite{Agrawal07}
\begin{equation}
\begin{array}{ll}
\partial_{z} U_{1} = &j \left[ \gamma_{1111}\left|U_{1}\right|^2 + 2 \gamma_{1133}\left|U_{3}\right|^2\right]U_{1}\\
& +j \gamma_{1113} U_{3} U_{1}^{*2}e^{j\Delta\phi\left(z\right) },\vspace{3mm}\\
\partial_{z} U_{3} =& j \left[ \gamma_{3333}\left|U_{3}\right|^2 + 2 \gamma_{3311}\left|U_{1}\right|^2\right]U_{3}\\
&+j \dfrac{1}{3}\gamma_{3111} U_{1}^{3}e^{-j\Delta\phi\left(z\right) },
\end{array}
\end{equation}
where $U_1$ and $U_3$ are the complex envelopes of the fundamental and TH waves, respectively, $\Delta\phi=\int_{0}^{z}\Delta\beta\left(z'\right) dz'$ is the phase mismatching, $\Delta\beta =\beta_3-3\beta_1  $ is the propagation-constant mismatching, $\beta_i=n\left(\omega_i\right) \omega_i /c$, $\omega$ is the angular frequency,   $n$ is the linear refractive index, $c$ is the speed of light, $\gamma_{ijkl}=n_2\omega_i/c A_{\mathrm{eff}}^{\left(ijkl\right)}$,  $n_2$ is the nonlinear refractive index in units m$^2$/W,   $A_{\mathrm{eff}}^{\left(ijkl\right)}$ is the effective area \cite{Stolen82},
\begin{equation}
A_{\mathrm{eff}}^{\left(ijkl\right)}=\frac{\sqrt{\prod_{u=i,j,k,l}\int\int dxdy\,\psi_u^2\left(x,y\right)  }}{\int\int dxdy\prod_{u=i,j,k,l}\psi_u\left(x,y\right)  },
\label{eq2}
\end{equation}
$x$ and $y$ are the transverse coordinates, and $\psi_i$ is the transverse profile of wave $i$.  (iv) Solve the differential equations in each segement in the regime of undepleted-pump approximation, the conversion efficiency of the TH can be written as 
\begin{equation}
\Gamma_{\mathrm{THG}}\left(z\right)= \dfrac{1}{9}P_{i}^2\left|\displaystyle\int_0^z dz\;\gamma_{3111}\left(z\right) \, \,e^{\,j\left[3 \gamma_{1111}P_iz-\Delta\phi\left(z\right) \right]}\right| ^2,
\label{eq3}
\end{equation}
with $P_i=\left| U_1 \left(0\right)\right|^2$ is the input power. If $\Delta\beta$ \textit{were} spatial-independent, a periodic modulation of $\gamma_{3111}$ could be used directly to correct the phase mismatching similar to QPM techniques in $\chi^{(2)}-$nonlinear media. It is worth to note that in tapered waveguides the phase mismatching   $\Delta\phi$ at any waveguide segment is related to the preceding segments, because of the integration that always runs from the beginning of the waveguide  \cite{Love91}. Assuming plane-waves     solutions in tapered waveguides could lead to pitfalls since $\Delta\beta$ varies along the structure. To measure the enhancement of the conversion efficiency using this technique,  the quantity
\begin{equation}
\Upsilon=\Gamma_{\mathrm{THG}}/\Gamma_{\mathrm{THG}}^{\mathrm{UNI}},
\end{equation}
is introduced, where $\Gamma_{\mathrm{THG}}^{\mathrm{UNI}}$ is the maximum conversion efficiency obtained using a uniform waveguide with an average transverse dimensions.

\begin{figure}
\centerline{\includegraphics[width=8.6cm]{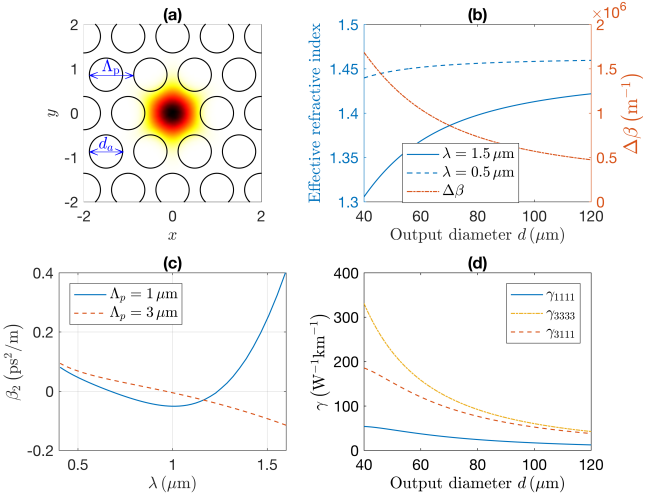}}
\caption{(Color online). Linear and nonlinear properties of a silica solid-core DOF  with $\Lambda_\mathrm{p}$ varies between 1 and 3 $\mu$m, a hole-diameter $d_a=0.75 \Lambda_\mathrm{p}$,  an output-diameter $d=40 \Lambda_\mathrm{p}$, and $n_2=2.25\times 10^{-20}$ m$^{2}$/W. The wavelength $\lambda$ of the pump and TH waves are 1.5 and 0.5 $\mu$m. (a) The fundamental-mode transverse profile of the TH at $\Lambda_\mathrm{p}=$ 1 $\mu$m. (b) The dependence of the effective refractive indexes and propagation-constant mismatching $\Delta\beta$ on the output diameter. (c) The wavelength dependence of the second-order dispersion coefficient $\beta_2$. (d) The dependence of the nonlinear coefficients $\gamma$  on the output diameter. These parameters will be used in the subsequent simulations presented in this paper, unless stated otherwise.
\label{Fig1}}
\end{figure}

Without loss of generality, I have used in simulations the parameters of dispersion oscillating fibres (DOFs), which are silica solid-core sinusoidally-tapered microstructured fibres, exploited in controlling modulation instability  \cite{Droques12,Droques13,Nodari15,Copie15}.  The fibre output diameter $d$ is given by
\begin{equation}
d(z)=d_{\mathrm{av}}\left[1-\Delta d\,\cos\left( 2\pi z/\Lambda_\mathrm{T}\right)\right],
\end{equation}
where $d_{\mathrm{av}}$ is the average diameter, and  $\Delta d$ is the amplitude of modulation. The linear and nonlinear  properties of a DOF with an outer diameter $d$ that  varies sinusoidally with $d_{\mathrm{av}}=80 \, \mu$m, and $\Delta d=0.5$ is shown in Fig. 1. The fibre is made of a stack of hollow capillary tubes with a pitch $\Lambda_\mathrm{p}$ varies between 1 and 3 $\mu$m and a hole-diameter $d_a=0.75 \Lambda_\mathrm{p}$. The mode profile of the TH at the smallest diameter is displayed in Fig. \ref{Fig1}(a), assuming a pump source with wavelength $\lambda= 1.5 \,\mu$m. The dependency of the effective refractive index of the fundamental and TH on $d$, and the corresponding $\Delta\beta$ are shown in panel (b). The simulations are performed using  `COMSOL', a commercial finite-element method,  including material dispersion of silica \cite{Saleh07}. The maximum fibre guiding loss is less than 0.1 dB/m.  The fibre second-order dispersion coefficient $\beta_2$ dependence on the wavelength at the minimum and maximum of $\Lambda_\mathrm{p}$ is displayed in panel (c). The TH wave is always in the normal dispersion regime, while the fundamental wave shifts continuously along the fibre between the normal and anomalous dispersion regimes. Figure \ref{Fig1}(d) presents the dependencies of some of the nonlinear coefficients $\gamma_{ijkl}$ on $d$.  The strong modulation of $d$ results in variation of $\gamma_{3111}$ between 37.8 and 185.5 W$^{-1}$km$^{-1}$. The nonlinear phase contribution to TH, $3 \gamma_{1111}P_i$, in  Eq. (\ref{eq3}) is safely neglected in this work, since  it requires at least a mega-watt CW pump source to be comparable to $\Delta\beta$. I have used in the simulations presented in the paper the parameters of this fibre, however, with different $\Delta d$. Numerical integrations should be performed with step size much smaller than the tapering period to avoid accumulation of computational errors. For a Gaussian pulse with 1 ps temporal duration and 1 W input power, modulation instability takes place roughly after more than 10 metres of propagation \cite{Agrawal07}.

\section{Quasi-phase-matching mechanism in  sinusoidally-tapered waveguides}
\begin{figure}
\centerline{\includegraphics[width=8.6cm]{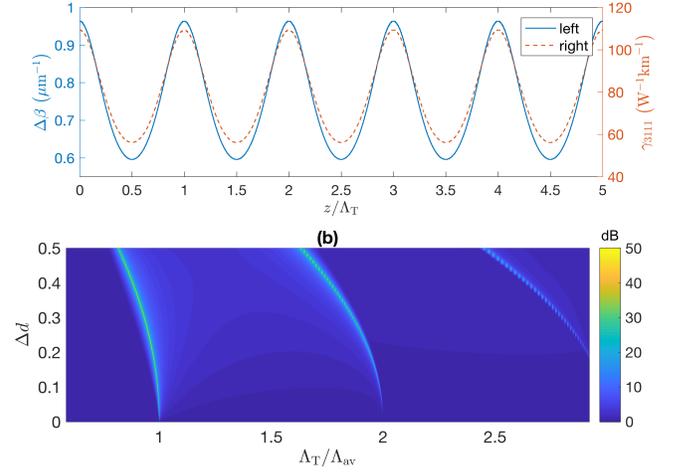}} 
\caption{(Color online). (a) Spatial dependence of $\Delta\beta$ and $\gamma_{3111}$ in a DOF used in Fig. \ref{Fig1} with $d_\mathrm{av}$ = 80 $\mu$m and  $\Delta d=0.2$. (b) Dependence of $\Upsilon$ on the tapering period $\Lambda_\mathrm{T} $ and modulation amplitude $\Delta d$ after propagating 1000 periods, with  $\Lambda_\mathrm{av}=8.56$ $\mu$m.
\label{Fig2}}
\end{figure}

Unlike QPM  in $\chi^{(2)}-$nonlinear media, both the nonlinear coefficient $\gamma_{3111}$ and the propagation-constant mismatching $\Delta\beta$ are periodically varying in a DOF, as depicted in Fig. \ref{Fig2}(a). $\Delta\beta\left(z\right)=\Delta\beta_m+\Delta\beta_w\left(z\right)$, where $\Delta\beta_m$   and  $\Delta\beta_w$ are  the material  and waveguide mismatchings, respectively.  $\Delta\beta_w$ has its strongest (weakest) value at the smallest (largest) fibre diameter. Hence, $\Delta\beta$ can be approximately written as 
\begin{equation}
\Delta\beta=\Delta\beta_{\mathrm{av}}\left[1+\Delta d\,\cos\left(\Omega_\mathrm{T} z\right)\right], 
\end{equation}
where  $\Delta\beta_\mathrm{av}$ is the average value of $\Delta\beta$, and $\Omega_\mathrm{T}=2\pi/\Lambda_\mathrm{T}$ is the tapering spatial frequency. There is a range of values of $\Delta\beta$ for a single nonlinear parametric process,  because of the periodic modulation of the fibre cross-section area. The corresponding phase mismatching can be written  in analog to uniform waveguides as $\Delta\phi=\widetilde{\Delta\beta}z$,  $\widetilde{\Delta\beta}=\Delta\beta_{\mathrm{av}}\left[1+\Delta d\,\mathrm{sinc}\left(\Omega_\mathrm{T} z\right)\right]$, and  $\mathrm{sinc}\left(x\right)=\sin\left(x\right)/x$. After a long propagation, $\widetilde{\Delta\beta}$ tends to $\Delta\beta_{\mathrm{av}}$. 

The novelty of this research is in obtaining the right QPM conditions that allow a DOF (or in general a PTW)    to be  employed as an efficient $\chi^{(3)}-$QPM structure. Figure \ref{Fig2}(b) shows the right combination of the tapering period and the amplitude of modulation that leads to strong enhancement ($\lesssim$ 50 dB) of TH generation  in comparison to a uniform waveguide with the same length. The tapering period is normalised to $\Lambda_\mathrm{av}=2\pi/\Delta\beta_\mathrm{av}$. The right $\Lambda_\mathrm{T}$ is  close to $\Lambda_\mathrm{av}$, however, it starts to deviate as $\Delta d$ increases. This matches with the sinc-behaviour of $\widetilde{\Delta\beta}$. In PTWs, the harmonics of  the nonlinear coefficients will be used to approximately correct $\Delta\beta_\mathrm{av}$, due to partial cancellation of the phase mismatching accumulated during the positive-half of the period with that during the negative-half. The bright trajectories shown in the plot corresponds to the fundamental period and its multiples. Interestingly, even for small modulation 1 -- 2 \%, $\Upsilon$ or $\Gamma_{\mathrm{THG}}$  could be dramatically enhanced.  For higher multiples, the enhancement drops and a stronger amplitude modulation is needed.  $\Delta\beta_\mathrm{av}$ is large in the range of $10^6$ for TH, which suggests a tapering period in the micrometer regime. Using  Fig. \ref{Fig1}(b), the condition  $\left|\partial_{z} \beta_{i}\right|  \ll \beta^2_{i}$ is still satisfied by approximately 2 order of magnitudes for this range of  tapering periods. 


\begin{figure}
\centerline{\includegraphics[width=8.6cm]{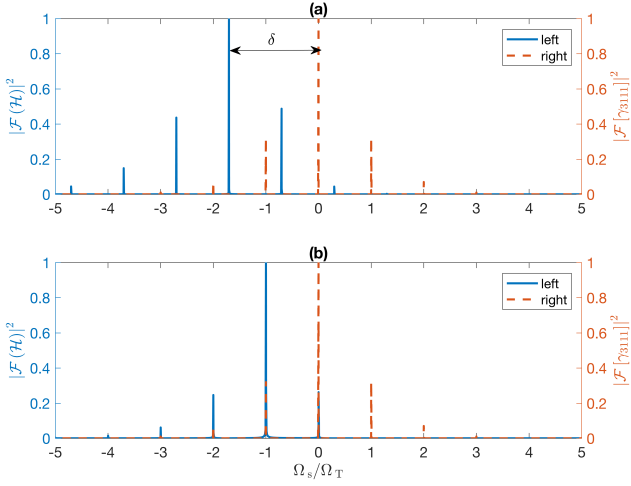}} 
\caption{(Color online). Power spectral densities of $ \mathcal{H}$ and $\gamma_{3111}$ in a DOF with $d_\mathrm{av}$ = 80 $\mu$m and $\Delta d=0.4$, (a) $\Lambda_\mathrm{T}/\Lambda_\mathrm{av} =1.5$, (b) $\Lambda_\mathrm{T}/\Lambda_\mathrm{av} =0.88$.  $\mathcal{F}$ is the Fourier-transform operator.
\label{Fig3}}
\end{figure}

The power-spectral densities of $\gamma_{3111}$ and $\mathcal{H}=\exp\left[-i\Delta\phi \right]$ are plotted vs the  spatial frequency $\Omega_\mathrm{s} $, normalised to $\Omega_\mathrm{T}$, in Fig. \ref{Fig3}(a,b). The Fourier spectrum  of $\gamma_{3111}$ is symmetric with a zero-component and multiples of $\Omega_\mathrm{T}$. Interestingly, I found that the Fourier spectrum of $\mathcal{H}$ is similar to that $\gamma_{3111}$, however, it is shifted to the left with multiples of another  spatial frequency $\Omega'_\mathrm{T}$. The strength of the sidebands of both spectra   increases  with $\Delta d$ and are  rapidly-decaying   for higher-order components. Hence, the integrand  in Eq. (\ref{eq3}) can be written using the Fourier series as
\begin{equation}
\gamma_{3111} \mathcal{H}=\displaystyle\sum_p\tilde{\gamma}_p e^{jp\Omega_\mathrm{T}z} \sum_q\tilde{h}_q e^{jq \Omega _\mathrm{T} z- j\delta z}\;,
\label{eq6}
\end{equation}
where $\delta=\Omega _\mathrm{T} -\Omega '_\mathrm{T} $ depends on the tapering period and the amplitude of modulation, $\tilde{\gamma}_p$ and  $\tilde{h}_q$ are the Fourier coefficients of $\gamma_{3111}$ and $ \mathcal{H}$, respectively, and $p,q=0,\pm1,\pm2,...$. When $\delta$ is zero or  $m\Omega _\mathrm{T}$ with $m$ an integer, the power spectral densities of $\gamma_{3111}$  and $\mathcal{H}$ will  coincide, as depicted in Fig. \ref{Fig3}(b). 

In periodically-poled waveguides, there is a single phase-mismatching Fourier component that will be counteracted by one of the harmonics of the periodic nonlinear coefficient by choosing the right poling period. Whereas in PTWs, there are multiples of phase mismatchings components with spatial frequencies $q \Omega _\mathrm{T} - \delta$. However, if $\delta=m\Omega _\mathrm{T}$, Eq. (\ref{eq6}) can rewritten as 
\begin{equation}
\gamma_{3111} \mathcal{H}=\displaystyle\sum_p\tilde{\gamma}_p e^{jp\Omega_\mathrm{T}z} \sum_{q'}\tilde{h}_{q'} e^{j q' \Omega _\mathrm{T} z}\;,
\label{eq8}
\end{equation}
with $q'=q-m$. In this case,   multiple opposite combinations of $\left(p,q'\right)$ such as $\left(0,0\right)$,$\left(1,-1\right)$, $\left(-1,1\right)$, $\left(-2,2\right)$, $\left(2,-2\right)$,... will be exploited simultaneously to correct the phase mismatching introduced by the periodic nature of the waveguide. A Fourier component of $\gamma_{3111}$ will  balance a component of  $ \mathcal{H}$ that has an opposite spatial frequency. Hence, the QPM condition for these combinations is
\begin{equation}
\mathrm{G}_{p,q'}=p\Omega_\mathrm{T}+q' \Omega _\mathrm{T}=0 .
\label{eq9}
\end{equation}
By proper adjustment of both $\Lambda_\mathrm{T} $ and $\Delta d$, this condition is satisfied and dramatic enhancement of TH generation is achieved as displayed in Fig. \ref{Fig2}(b). Deviating from the perfect QPM condition of the PTWs results in deteriorating the conversion efficiency.  The spectrum $\left|\mathcal{F}\left(\mathcal{H}\right)\right|^2$ continues to shift to the left as  $\Lambda_\mathrm{T} $ increases. Hence, $\Upsilon$ or $\Gamma_{\mathrm{THG}}$ starts to  drop for higher values of $\Lambda_\mathrm{T} $ due to a weakly overlap between the strong spectral components of $\gamma_{3111}$ and $\mathcal{H}$.  This can be recovered via using longer waveguides,   materials with large effective nonlinearity, or large input power.

\begin{figure}
\centerline{\includegraphics[width=8.6cm]{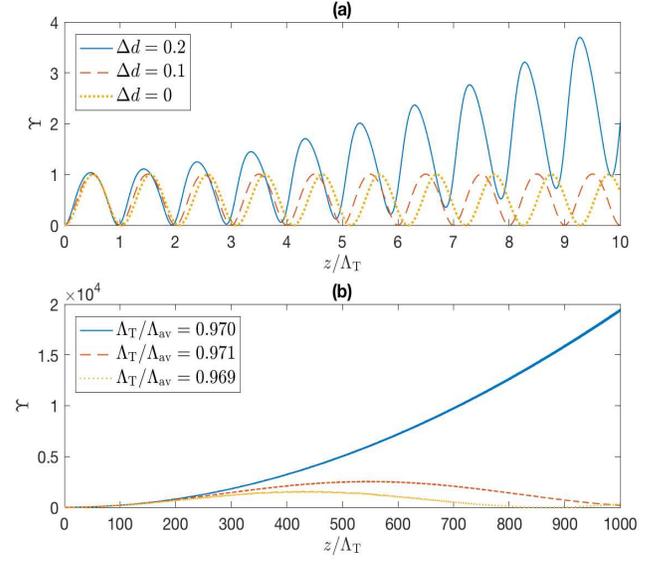}} 
\caption{(Color online). Spatial dependence of $\Upsilon$  in a DOF with $d_\mathrm{av}$ = 80 $\mu$m, and $P_i=1$ W. (a) Short scale with  $\Delta d=0,0.1,0.2$, and  $\Lambda_\mathrm{T}/\Lambda_\mathrm{av} =0.97$. (b) Long scale with  $\Delta d=0.2$, and  $\Lambda_\mathrm{T}/\Lambda_\mathrm{av} =0.97,0.971, 0.969$. 
\label{Fig4}}
\end{figure}

The conversion efficiency  can be rewritten as 
\begin{equation}
\Gamma_{\mathrm{THG}}\left(z\right)= \dfrac{1}{9}P_{i}^2z^2\left|\displaystyle\sum_{p,q'}\tilde{\gamma}_p \tilde{h}_{q'}\mathrm{sinc}\left(\mathrm{G}_{p,q'}z/2\right)e^{j\mathrm{G}_{p,q'}z/2}\right| ^2,
\label{eq10}
\end{equation}
using Eq. (\ref{eq8}). Therefore, $\Gamma_{\mathrm{THG}}$ depends on a superposition of multiple sinc functions, modulated with different complex amplitudes and relative phases. Panels (a,b) in  Fig. \ref{Fig4} depicts the spatial dependence of $\Upsilon$ over short  and long propagation distances. For comparison, the spatial dependence of $\Upsilon$ in a uniform waveguide and when $\Delta d$ does not equal to the right value  are also shown in Fig. \ref{Fig4}(a). Opposite combinations with $p+q'=0$, leads to a quadratic spatial dependence and the growth of $\Gamma_{\mathrm{THG}}$. Contrarily, other combinations that do not satisfy the QPM condition,  Eq. (\ref{eq9}), such as $\left(0,-1\right)$,$\left(0,1\right)$, $\left(0,-2\right)$, $\left(-2,1\right)$,..., results in a sinusoidal behaviour and does not contribute to the growth of the the conversion efficiency. The strongest dominant component of these combinations is $\mathrm{G}_{0,-1}$, as shown in Fig. \ref{Fig3}(b).

On a macroscopic scale, $z\ll  1/\left|\mathrm{G}_{0,-1}\right|$, $\Gamma_{\mathrm{THG}}$ has the classical quadratic spatial dependence. Afterwards, $\Gamma_{\mathrm{THG}}$ starts to sinusoidally oscillates and approaches zero on an intermediate scale,  when $z$ is multiples of $2\pi/\left|\mathrm{G}_{0,-1}\right|$.   An overall gain can be obtained after long propagation  if the amplification due to the $\left(p,q'\right)$ opposite combinations  overcomes the sinusoidal oscillation due to $\mathrm{G}_{0,-1}$ component within each period, as displayed in Fig. \ref{Fig4}(b) [solid-blue curve]. The result would behave as an amplified sinusoidal wave.

In PTWs, there is  a coherence length $l_{c}^{(s)}=2\pi/\left|\mathrm{G}_{0,-1}\right|$ that corresponds to the short-scale sinusoidal oscillations. There is also another long-scale coherence length $l_{c}^{(l)} $ due to the non-ideal compensation of the phase mismatching, $l_{c}^{(l)}=2\pi/\left|\mathrm{G}_{p,q'}\right|$, with $\left(p,q'\right)=\left(0,0\right)$ or $\left(1,-1\right)$. These are the two strongest Fourier components that lead to the growth of the TH. For perfect alignment between the spectra of $\gamma_{3111}$ and $ \mathcal{H}$, $\mathrm{G}_{0,0}=\mathrm{G}_{-1,1}=0$, and $l_{c}^{(l)}$ goes to infinity. As shown in Fig. \ref{Fig4}(b), $\Upsilon$ is reduced by approximately an order of magnitude by slight change of $\Lambda_\mathrm{T}$. For the first  200 -- 300 periods of propagation, $\Upsilon$ is approximately the same for the three values of $\Lambda_\mathrm{T}$. Afterwards, it decays  for the non-ideal values of $\Lambda_\mathrm{T}$ due to the accumulation of the phase-mismatching. Fine tuning is necessary for long propagation, similar to periodic poling technique \cite{Fejer92}, since the linewidth of the Fourier components of $\gamma_{3111}$ and $ \mathcal{H}$ becomes very sharp as  the structure length increases. This restricts the values of $\Lambda_\mathrm{T}$ and   $\Delta d$ to allow an alignment between the two spectra. The fine tuning is relaxed for higher values $\Delta d$ as depicted  by a broadening of the bright trajectories in Fig. \ref{Fig2}(b). 

\begin{figure}
\centerline{\includegraphics[width=8.6cm]{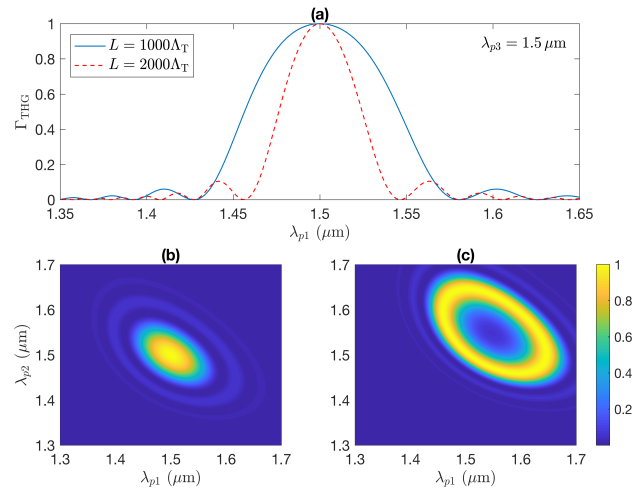}} 
\caption{(Color online). Wavelength-dependence of the normalised conversion efficiency of the TH generation  at the end of a DOF with   $d_\mathrm{av}$ = 80 $\mu$m, $\Delta d=0.2$, $\Lambda_\mathrm{T}/\Lambda_\mathrm{av} =0.97$, and $P_i=1$ W. $\lambda_{p1}$, $\lambda_{p2}$, $\lambda_{p3}$ are the three pump photons at and around the fundamental frequency 1.5 $\mu$m. (a) 1D representation with the TH at 500 nm. (b,c) 2D representation with the TH at 500 and 516 nm, respectively, and $ L = 1000 \Lambda_\mathrm{T}$. 
\label{Fig5}}
\end{figure}

The dependence of the conversion efficiency on the pump wavelength at the end of a waveguide with length $L$ is depicted in Fig. \ref{Fig5}. Panel (a) displays the case, in which one pump photon frequency is fixed at 1.5 $\mu$m, whereas the frequencies of the other two photons are scanned around that value, such that the TH is  0.5 $\mu$m. 
Using Eq. (\ref{eq10}), only terms with  $\mathrm{G}_{p,q'}\approx 0$ will contribute to the output spectrum at a fixed length. Hence, $\Gamma_{\mathrm{THG}}$ will behave as a squared sinc-function with weak sidelobes, and spectral bandwidth that is inversely proportional to $L$, as demonstrated in Fig. \ref{Fig5}(a). Panels (b,c) show the 2D representation of $\Gamma_{\mathrm{THG}}$ (equivalent to the phase-matching function of  photon triplets \cite{Cavanna16,Moebius16}) over which the energy and momentum conservation conditions are satisfied for  TH at 500 and 516 nm, respectively.  Because of the weakness of the sidelobes,  the spectral purity of  three downconverted photons out of one pump photon (reverse process of TH generation) is anticipated to be  high. The spectrum maintains the feature of having a single mainlobe over a relatively broad range of a TH between 500$\pm$10 nm.

\section{Discussions and conclusions}
Using  tapered single-mode waveguides can lead to a power loss due to coupling of the fundamental mode with higher-order cladding modes. This would impose an additional condition on the acceptable values of  $\Delta d$ and  $\Lambda_\mathrm{T}$ that can ensure an \textit{adiabatic propagation} inside PTWs with minimal losses. Using the weak-power-transfer criterion \cite{Love91,Yerolatsitis14}, the  condition that delineates between adiabatic and lossy tapering regimes can be written as,
\begin{equation}
\alpha=\left|\frac{\sigma\left(z\right)\lambda}{2n_\mathrm{max}\left[n_a^2\left(z\right)-n_b^2\left(z\right)\right]}\frac{dr}{dz}\right|\ll 1,
\label{eq11}
\end{equation}
where $n_\mathrm{max}$ is the maximum core refractive index, $a,b$ refers to the fundamental mode and the closest higher-order cladding mode, $n_i$ is the local effective refractive index of mode $i$, $\sigma$ is the overlap integral of the two mode profiles with the radial-derivative of  the refractive index transverse distribution,
\begin{equation}
\sigma\left(z\right)=\frac{\tilde\sigma\left(z\right)}{r_\mathrm{av}}=\frac{1}{r_\mathrm{av}}\frac{\displaystyle\int\int d\rho d\theta\,\psi_a\psi_b \,\frac{dn^2}{d\rho}}{\sqrt{\displaystyle\int\int d\rho d\theta\,\psi_a^2 \displaystyle\int\int d\rho d\theta\,\psi_b^2 }},
\end{equation}
$\tilde\sigma$ is the normalised overlap integral, $r_\mathrm{av}$ is the average fibre radius,   the field distribution $\psi$ is defined in the polar coordinates $r$  and $\theta$,   $r$ is the  radial distance,  $\theta$ is the azimuthal angle, and $\rho=r/r_\mathrm{av}$. For PTWs, substituting $r=r_\mathrm{av}\left[1-\Delta d\,\cos\left( 2\pi z/\Lambda_\mathrm{T}\right)\right]$,  Eq. (\ref{eq11}) becomes
\begin{equation}
\alpha=\left|\frac{\pi\tilde{\sigma}\lambda\Delta d\,\sin\left( 2\pi z/\Lambda_\mathrm{T}\right)}{n_\mathrm{max}\Lambda_\mathrm{T}\left(n_a^2-n_b^2\right)}\right|\ll 1.
\end{equation}
Hence, small amplitude of modulation and large tapering period is required to allow for adiabatic propagation, as intuitively expected.  A full analysis of the above criterion requires numerical calculations of all the modes of the system along a length of at least a half tapering period with very small increments. This would be a subject of future studies. However,  estimated values of $\alpha$ could be determined for the proposed structure of TH generation.  Both $\tilde{\sigma}$ and $n_a^2-n_b^2$ slightly change  along the waveguide. Hence, the adiabatic criterion will depend on the sine term that reaches its maximum unity when $z$ is odd multiple numbers of $\Lambda_\mathrm{T}/4$. This corresponds to the positions where the fibre radius equals to its average value. The estimated maximum   value of $\alpha$ at these positions is $\approx0.2$ assuming  $\lambda=0.5\,\mu$m, $\tilde{\sigma}=0.5$,  $n_\mathrm{max}=1.45$, $n_a=1.44$, and $n_b=1.43$, $\Delta d =0.1$, and $\Lambda_\mathrm{T}=8.5\,\mu$m. So,  the criterion is moderately satisfied  at and around these positions. However, when $z$ is multiples of $\Lambda_\mathrm{T}/2$, the criterion is completely satisfied. Therefore, to avoid accumulation of  non-adiabatic losses, a structure with a smaller  $\Delta d $ [as shown in Fig. \ref{Fig2}(b)], suppressed lossy higher-order cladding modes, or a careful designed waveguide length should be used.

The possible longest  tapering period required to obtain high  $\Gamma_{\mathrm{THG}}$ is approximately few tens of microns using Fig. \ref{Fig2}(b),  which is currently unaccessible via the current DOF fabrication technology. Currently, the  state-of-the-art of DOF fabrication is limited to a period of few tens of centimetres over about 200 m-long \cite{Nodari15,Copie15}. However with the rapid progress in the fabrication methods and via using advanced post-treatment processes,  I would imagine that  this limitation will be mitigated in the  future. Alternatively, these estimated tapering periods could be realised sooner using other waveguide platforms such as  laser-written and  planar waveguides  \cite{Davis96,Belt17}, which are  more promising techniques and will be considered in future investigations as potential candidates for experimental demonstration.  In fact, width-modulated sinusoidal tapering with 1 mm tapering period has been demonstrated in high-nonlinear rectangular silicon nanowires over 5-mm long using e-beam lithography \cite{Driscoll12}.  Also in these platforms, temperature-tuning effect  \cite{Jasny04,Leviton06} can be utilised to correct any phase-mismatching introduced by fabrication tolerance via integrating thermo-optic switches. $\chi^{(2)}-$QPM periodically-poled microstructures were demonstrated after two decades of their proposal. So, one could anticipate that on-demand PTWs will be available within the same period, or may be even less.



In conclusion, I have theoretically demonstrated  the concept of having efficient QPM schemes in third-order nonlinear materials using periodically-tapered waveguides. In this paper, I have studied as an example enhancing third-harmonic generation in longitudinally sinusoidally tapered fibres. However, the study is applicable for other FWM  processes such as parametric amplification, where the phase mismatching is relatively arbitrary and the corresponding tapering period  can be set by the current fabrication technologies. In these structures, there are multiple values of the  phase mismatching of a single nonlinear process due to the periodic nature of the waveguides.  However, the QPM condition is satisfied by using a right combination of the tapering period and modulation amplitude that allow an alignment between the Fourier spectra of  the nonlinear coefficient and the phase-mismatching term.  Simulations show an enhancement of the TH conversion efficiency by 50 dB after 1000 periods of propagation. Finally, I envisage that this work will  stimulate and open new areas of research.  In fact, it could revolutionise  the applications of integrated nonlinear optics using  materials, such as silicon or other CMOS-compatible compounds that are incompatible with periodically-poled techniques.

\paragraph*{Acknowledgment.} The author would like to thank Prof. A. Mussot and Dr. A. Kudlinski in University of Lille in France, as well as Dr. F. Biancalana, Dr. A. Fedrizzi and Prof. R. Thomson in Heriot-Watt University in UK for useful discussions. The author would like also to acknowledge the support of his research by Royal Society of Edinburgh (RSE).


\bibliographystyle{apsrev4-1}	

%

\end{document}